\documentstyle[12pt]{article}
\begin{document}
\def\be{\begin{equation}}
\def\ee{\end{equation}}
\def\ba{\begin{eqnarray}} 
\def\ea{\end{eqnarray}}

\newcommand{\bbf}{\mathbf}
\newcommand{\rrm}{\mathrm}

\title{ Localization Properties of Two Interacting Electrons in a 
Disordered Quasi One--Dimensional Potential\\}

\author{Jean Richert\\
Laboratoire de Physique Th\'eorique\footnotemark[1],\\
Universit\'e Louis Pasteur, 67084 Strasbourg Cedex, France 
\\
and \\
Hans A. Weidenm\"uller\\
Max-Planck-Institut f\"ur Kernphysik,
D-69029 Heidelberg, \\
Germany}

\date{\today}
\maketitle
\begin{abstract} 
We study the transport properties of two electrons in a quasi
one--dimensional disordered wire. The electrons are subject to both,
a disorder potential and a short range two--body interaction. Using
the approach developed by Iida {\it et al.} [ Ann. Phys. (N.Y.)
{\bf 200} (1990) 219 ], the supersymmetry technique, and a suitable
truncation of Hilbert space, we work out the two--point
correlation function in the framework of a non--linear $\sigma$ model.
We study the loop corrections to arbitrary order. We obtain a
remarkably simple and physically transparent expression for the change
of the localization length caused by the two-body interaction.
\end{abstract}

\footnotetext[1]{UMR 7085, ULP/CNRS}    
Non--interacting electrons in a one--dimensional random potential are
localized. The presence of a short--range two--body interaction (a
screened Coulomb potential) affects the localization length. For two
interacting electrons in one dimension, this fact has been clearly
established in a series of numerical and analytical investigations
(see Refs.~\cite{dima},\cite{imry},\cite{song} and references therein).
The localization length increases independently of the sign of the
interaction when the electrons move together at short
distance~\cite{song}.

In the present work we address the more realistic and more demanding
case of two electrons in a quasi one--dimensional disordered wire
interacting via a short--range interaction. Using an analytic method
developed recently~\cite{ben}, a reduction of Hilbert space and
supersymmetry, we achieve a complete analytical solution of the
problem. We show that the two--body interaction affects the
localization length via the level density. We establish a criterion
for the onset of interaction--induced effects on the localization
length. Our final analytical expression allows for the numerical
determination of the localization length for any given two--body
interaction. To the best of our knowledge, this is the first time
that an analytical solution of a problem involving both, disorder and
interaction, has been achieved.

{\it Model}. We consider a quasi one--dimensional wire of length $L$
which consists~\cite{iid} of $K$ slices labelled $a,b,c,\ldots,K$.
The surfaces connecting neighboring slices are transverse to the axis
of the wire. Eventually we take the limit where $K \rightarrow \infty$
and where the longitudinal extension of the slices goes to zero. In
each slice we use an arbitrary basis of single--particle states
labelled $| a j \rangle$ with $j = 1,2, \ldots, l$ and Fermionic
creation and annihilation operators $\alpha^{\dagger}_{aj}$ and
$\alpha_{aj}$, respectively. We later take the limit $l \rightarrow
\infty$.

The Hamiltonian $H$ is the sum of three terms, $H = H_0 + H_1 + H_2$.
In every slice, disorder is simulated by an ensemble of
random single--particle Hamiltonians belonging to the unitary ensemble, 
\be
H_0 = \sum_{aij} h^{(a)}_{ij} \alpha^{\dagger}_{ai} \alpha_{aj}
\label{eq0} \ .
\ee
The complex random variables $h^{(a)}_{ij}$ have a Gaussian
distribution with mean value zero and the following non--vanishing
second moments,
\be
\overline{h^{(a)}_{ij} h^{(a')*}_{i'j'}} = \frac{\lambda^2}{l}
\delta_{a a'} \delta_{i i'} \delta_{j j'} \ .
\label{eq1}
\ee
The overbar denotes the ensemble average and $\lambda$ has the
dimension of energy. The disorder Hamiltonians in different slices are
uncorrelated. Hopping between neighboring slices is described by the
hopping term
\be
H_{1} = \sum_{a i} v [ \alpha^{\dagger}_{ai} \alpha_{a+1 i} + h.c.]
\ .
\label{eq2}
\ee
The short--range two--body interaction acts only between electrons
in the same slice and reads
\be
H_2 = \sum_a \sum_{i<j,i'<j'} w_{iji'j'} \alpha^{\dagger}_{ai}
\alpha^{\dagger}_{aj} \alpha_{aj'} \alpha_{ai'} \ .
\label{eq4}
\ee
The two--body matrix elements are antisymmetric in the pairs $(ij)$
and $(i'j')$ and Hermitean. Both $v$ and the $w_{iji'j'}$'s are
non--random quantities independent of the slice label $a$. For the
single--electron problem, it has been shown via
supersymmetry~\cite{iid} that $H_0 + H_1$ defines the same
non--linear sigma model as obtained from a continuum model using a
kinetic term and a random potential. This is why our model is generic.

The Hilbert space of the two--electron problem is spanned by the
orthonormal states $|a b \mu \rangle$ where $a \leq b$ and where $\mu$
stands for the pair $(i,j)$ with $i < j$ for $a = b$. For $a < b$
fixed, the number $N_{ab}$ of states is $l^2$ while for $a = b$,
$N_{aa} = l(l-1)/2$. The totality of states $|a b \mu \rangle$ with
$\mu = 1, \ldots, N_{ab}$ is referred to as the box $(ab)$. The effect
of the two--body interaction on localization properties in the
one--dimensional case is strongest when the two electrons stay close
to each other~\cite{song}. This fact, and the need to reduce the
dimension of Hilbert space, cause us to consider only the $2K-1$ boxes
where the two electrons are either located in the same slice $a$ (this
is box $(aa)$), or in two adjacent slices $a$ and $a+1$ (this is box
$(a(a+1))$), with $a = 1, \ldots, K$.

{\it Supersymmetry}. Localization properties can be read off the
two--point correlation function
\be
C(n) = \overline{|\langle a (a+1) \mu| (E^+ - H)^{-1} | (a+n) (a+n+1)
\nu \rangle|^2} \ .
\label{a1}
\ee
To be independent of edge effects, we choose $1 \ll a \ll (a+n) \ll K$.
For large $n$, $C(n)$ should decay exponentially in $n$. In proper
units, the coefficient in the exponent defines the localization length.
We calculate $C(n)$ using supersymmetry~\cite{efe,ver}.

Because of the Gaussian distribution of the coefficients $h^{(a)}_{ij}$,
the ensemble average in Eq.~(\ref{a1}) is entirely determined by the
second moments of the matrix elements $\langle a b \mu | H_0 | a b \sigma
\rangle$ of $H_0$ with $b=a,(a+1)$, i.e., by the quantities
\be
{\cal A}^{(k)}_{\mu \nu ; \rho \sigma} = \frac{l}{\lambda^2}
\overline{ \langle a b \mu | H_0 | a b \sigma \rangle
\langle a' b' \rho | H_0 | a' b' \nu \rangle }
\label{eq5} \ .
\ee
The index $k$ labels the various possible combinations of $(a,b,a',b')$.
In the direct product space $\{\mu \nu\}$, the matrix ${\cal A}^{(k)}$
can be diagonalized and expanded into bilinear forms of its 
eigenvectors~\cite{ben},
\be
{\cal A}^{(k)}_{\mu \nu ; \rho \sigma} = \frac{1}{N(k)} \sum_{s = 0}^1
\Lambda^{(s)}(k) \sum_{\tau} C^{(s \tau)}_{\mu \nu}(k;l) C^{(s
  \tau)}_{\rho \sigma}(k;r)) \ ,
\label{eq6}
\ee
where the $\Lambda^{(s)}(k)$ are real positive eigenvalues, $C^{(s
\tau)}$ the corresponding right ($r$) and left ($l$) eigenvectors,
$\tau$ distinguishes degenerate eigenvectors, and $N(k) \gg 1$ is a 
suitably chosen normalization factor related to $N_{aa}$ and
$N_{a(a+1)}$.

The generating function $Z$ carries in the exponent terms of the form
\be
- \frac{i}{2} \sum_{a b} \sum_{\mu \nu} \Psi^{*}_{a b \mu} L^{1/2}
\langle a b \mu | H_0 | a b \nu \rangle L^{1/2} \Psi_{a b \nu} \ .
\label{eq7}
\ee
Here $\Psi_{a b \mu}$ is a graded vector. The graded matrix $L$ is
defined in Ref.~\cite{ver}. After averaging $Z$, the quartic terms
in the $\Psi$'s appearing in the exponent involve the matrices
${\cal A}^{(k)}$. The eigenvalue expansion~(\ref{eq6}) allows us to
use the Hubbard--Stratonovich (H--S) transformation. We introduce
the matrices
\ba
A^{( a b a' b'; \tau)}_{\alpha \beta}(p) &=& i \lambda \sum_{\mu \nu}
L^{1/2}_{\alpha \alpha} \Psi_{a b \nu \alpha} \Psi^{*}_{ a' b' \mu
\beta} L^{1/2}_{\beta \beta} \nonumber \\
&& \times C^{1 \tau}_{\mu \nu}( a b a' b'; p)
\label{eq8}
\ea
where $p$ specifies whether a right-- or left--hand eigenvector $C$ is
involved. The terms quartic in the $\Psi$'s are bilinear (rather than
quadratic) in the matrices $A$. This fact necessitates the introduction
of suitable linear combinations of the $A$'s. In the H--S tranformation,
each such linear combination generates a $\sigma$-field. After
integration over the $\Psi$'s, the ensemble average of $Z$ takes the
form
\ba
&&\overline{Z} = \int {\rm d} [\sigma] \exp \biggl [ - \sum_k {\rm
  trg} \biggl ( N(k) (\sigma^{(k)})^2 \nonumber \\  
&& - \ {\rm trg} \ {\rm tr} \ \ln \ {\bf N}(J) \biggr )
   \biggr ] \ .
\label{eq9}
\ea   
The sum over $k$ extends over all $\sigma$-fields, the volume element
is denoted by ${\rm d}[\sigma]$. Here ${\bf N}(J)$ is both, a graded
matrix and a matrix in  Hilbert space with basis vectors $|a b \mu
\rangle$ and $b = a,(a+1)$. The symbol $J$ denotes the source
terms~\cite{ver}. For lack of space, we give the structure of
${\bf N}(J)$ only for the cases $a=a',b=b'$ where
\ba
&&\langle a a \mu | {\bf N}(J) | \ a a \nu \rangle = \biggl (E -
\lambda \sigma^{(aa)} \biggr ) \delta_{\mu \nu} \nonumber
\\
&&\qquad - \sum_{\tau} \lambda \sigma^{(aa; \tau)} C^{(1
  \tau)}_{\mu \nu}(0) \ - w_{\mu \nu} \ 
\label{eq10}
\ea
and
\ba
&&\langle a (a+1) \mu | {\bf N}(J) | \ a (a+1) \nu \rangle = \biggl
(E - \nonumber \\
&&\qquad - \lambda \sigma^{(a (a+1))} \biggr ) \ \delta_{\mu \nu}
- \sum_{\tau} \lambda \sigma^{(a (a+1); \tau)} C^{(1 \tau)}_{\mu
\nu}(1) \ .
\label{eq11}
\ea
Non--diagonal elements like $\langle (a-1) a \mu | {\bf N}(J) | a a
\nu \rangle$ contain hopping contributions.

{\it Saddle Points}. The factors $N(k)$ in Eq.~(\ref{eq9}) obey $N(k)
\gg 1$. Hence, we use the saddle--point approximation with $v =
0$~\cite{iid}. The saddle--point equations can be cast into the form
of a Pastur equation, 
\ba
&&\langle a b \mu | X | a' b' \nu \rangle = \sum_{\rho \sigma} \\
&&\times \overline{ \langle a b \mu | H_0 | a b \sigma \rangle
\biggl (\frac{1}{E - w -X} \biggr )_{(a b|a' b'); \sigma \rho} \langle
a'b' \rho | H_0 | a' b' \nu \rangle } \ . \nonumber
\label{eq12}
\ea
The matrix $X$ contains all the $\sigma$--matrices which occur in
${\bf N}(J)$.

Eq.~(\ref{eq12}) implies that the solutions $\sigma^{(k)}_{\rm sp}$ are
zero or approximately zero except for those with $k=(a,(a+1),a,(a+1))$
and $k=(a,a,a,a)$. The corresponding fields obey the equations
\be
\sigma^{(a(a+1))}_{\rm sp} = \frac{\lambda}{E - \lambda
\sigma^{(a(a+1))}_{\rm sp}}
\label{eq13}
\ee
and 
\be
\sigma^{(aa)}_{\rm sp} = N_{aa}^{-1} \ {\rm tr} [ ( \frac{\lambda}{E
- w - \lambda \sigma^{(aa)}_{\rm sp}} )_{\mu \nu} ] \ .
\label{eq14}
\ee
The diagonal solution for $\sigma^{(a(a+1))}_{\rm sp}$ takes the
standard form $\sigma^{(a(a+1))}_{\rm d}$ $=$ $(E/(2 \lambda))$
$\pm i$ $\Delta_{1}(E)$, with $\Delta_{1}(E)$ $=$ $\sqrt{1 - (E/(2
\lambda))^2 }$.
Here $\Delta_{1}(E)$ is proportional to the spectral density
$\rho_{{\rm sp} 1}(E)$ in box $(a(a+1))$ and has the shape of
a semicircle. This fact points to a deficiency of the saddle--point
approximation: The two electrons in slices $a$ and $(a+1)$ do not
interact. The spectrum of each has semicircular shape. The spectrum
in box $(a(a+1))$ is, thus, the convolution of two semicircles and not
a semicircle. This shows that we must take into account loop
corrections to the saddle--point approximation. This is done below.
The invariance of the saddle--point equation Eq.~(\ref{eq13}) under
pseudounitary graded transformations $T$ implies that the saddle--point
manifold has the
form~\cite{ver}
\be
\sigma^{(a(a+1))}_{\rm sp} = T^{(a (a+1))} \ [ \ \frac{E}{2 \lambda}
- i \ \Delta_{1}(E) \ L \ ] \ ( T^{(a (a+1))} )^{-1} \ .
\label{eq17}
\ee
Eq.~(\ref{eq14}) for $\sigma^{(aa)}_{\rm sp}$ contains the two-body
interaction $w$. Diagonalizing $w$ and using a simple geometrical
construction shows that Eq.~(\ref{eq14}) possesses either $N_{aa}$ 
real solutions (this applies for $E < E_1$ and for $E > E_2$), or,
for $E_1 < E < E_2$, $(N_{aa} - 2)$ real and two complex conjugate
solutions $a(E) \pm i \Delta_0(E)$. Thus,
\be
\sigma^{(aa)}_{\rm sp} = T^{(aa)} \ [ \ a(E) - i \Delta_{0}(E) \ L
\ ] \ (T^{(aa)})^{-1} \ ,
\label{eq18}
\ee
where $\Delta_{0}(E) > 0$ is proportional to the spectral density
$\rho_{{\rm sp} 0}(E)$ in box $(aa)$. The essential difference
between the saddle--point solutions in Eqs.~(\ref{eq17}) and
(\ref{eq18}) lies in the difference between $\Delta_{0}(E)$ and
$\Delta_{1}(E)$, i.e, in the different spectral densities $\rho_{{\rm
    sp} 0}(E)$ and $\rho_{{\rm sp} 1}(E)$. The interaction $w$ deforms
the semi--circle $\Delta_{1}(E)$. Aside from a possible overall shift
of the spectrum, $w$ causes $\Delta_{0}(E)$ to be smaller in the
centre of the semicircle, and to extend beyond the end points $(-2
\lambda, +2 \lambda)$ of $\Delta_{1}(E)$. For a weak interaction, we
calculate $E_{1,2}$ using a second--order perturbation expansion in
$w$ which yields
\be
E_{1,2} = E_0 \pm (2 + U^2) \lambda \ .
\label{eq19}
\ee
Here $E_0 = (1/N_{aa}) {\rm tr} (w)$ and $U^2 = \{ (1/N_{aa}) {\rm tr}
(w^2) - [(1/N_{aa}) {\rm tr} (w)]^2 \} / \lambda^2$. This shows shift
and widening of the spectrum. If we identify the matrix elements
$h^{(a)}_{ij}$ with those of the impurity potential $V_{\rm imp}$, it
is easy to see that a qualitative change of the spectral density occurs
whenever
\be
\langle a a \mu | w^2 | a a \mu \rangle_{\rm av} \geq \langle a i |
(V_{\rm imp})^2 | a i \rangle_{\rm av}
\label{eq20}
\ee
where the matrix elements are averaged over $\mu$ and $i$, respectively.
This is the announced qualitative criterion.

{\it Localization}. To calculate localization properties, we expand the
effective Lagrangian in the exponent of Eq.~(\ref{eq9}) in powers of
$v$, keeping terms up to second order. We put $J = 0$ and
use that for $k = (a,a,a,a)$ and $k = (a,a+1,a,a+1)$, the
$\sigma^{(k)}_{\rm sp}$'s are solutions of the saddle--point
Eqs.~(\ref{eq13}) and (\ref{eq14}). We also use Eqs.~(\ref{eq17}) and
(\ref{eq18}). The terms quadratic in the $\sigma^{(k)}_{\rm sp}$'s
vanish, and the matrix ${\bf N}$ yields
\ba
&& + (v/\lambda)^2 (3 l^2 / 4 ) \Delta_{0} \Delta_{1} \sum_{j =
  1}^{K'} {\rm trg} \biggl ( T^{(j)} L (T^{(j)})^{-1}  \nonumber \\
&&  \times T^{(j+1)} L (T^{(j+1)})^{-1} \biggr )
\label{eq22}
\ea
where the $2K - 1$ boxes $(aa)$ and $a(a+1)$ of the two--electron
problem have been mapped onto $K' = 2K - 1$ slices by putting $j =
2a -1$ for the boxes $(aa)$ and $j = 2a$ for the boxes $(a(a+1)$.
The expression~(\ref{eq22}) has exactly the form of the non--linear
sigma model derived for the transport of a single electron through
a disordered wire in Ref.~\cite{iid}. Since $\Delta_{0}$ and
$\Delta_{1}$ are proportional to the spectral densities
$\rho_{{\rm sp} 0}(E)$ and $\rho_{{\rm sp} 1}(E)$ in boxes
$(aa)$ and $(a(a+1))$, respectively, as determined by the
saddle--point condition, Eq.~(\ref{eq22}) suggests defining the
saddle--point approximation $T_{\rm sp}$ to the transport coefficient
by
\be
T_{\rm sp} = 2 \pi \ \rho_{{\rm sp} 0}(E) \ v^2 \ \rho_{{\rm sp} 1}(E)
\ .
\label{eq23}
\ee
The localization length is proportional to $T_{\rm sp}$. We conclude
that the ratio of the localization lengths $\zeta(w \neq 0)$ for
non--vanishing two--body interaction and $\zeta(w = 0)$ for vanishing
two--body interaction is given by 
\be
\frac{\zeta(w \neq 0)}{\zeta(w = 0)} = \frac{\rho_{{\rm sp} 0}(E, w
  \neq 0)}{\rho_{{\rm sp} 0}(E, w = 0)} \ .  
\label{eq24}
\ee
This result shows that and how the two-body interaction modifies the
localization length. Depending upon the energy considered and on the
sizes of the shift and of the widening of the spectrum in box $(aa)$
compared to the unperturbed spectrum in box $(a(a+1))$, the
localization length may decrease or increase. In general, we expect
an increase near the edge of the unperturbed spectrum, and a decrease
in the center.

{\it Loop Corrections}. The result Eq.~(\ref{eq24}) is obtained from
the saddle--point approximation, and it was remarked earlier that
this approximation is deficient. Therefore, we now study the loop
corrections to the saddle--point solution. We do so by expanding the
$\sigma$'s around the saddle--point solutions $\sigma^{(k)}_{\rm sp}$.
We recall that some of these vanish. We write
\be
\sigma^{(k)} = \sigma^{(k)}_{\rm sp} + T^{(k)} \delta \sigma^{(k)}
(T^{(k)})^{-1} \ .
\ee
We expand $\overline{Z}$ in powers of the $\delta \sigma^{(k)}$'s,
keeping in the exponent only the quadratic terms which originate
from the corresponding terms in Eq.~(\ref{eq9}). We need to
address the form of the source terms $J$ contained in ${\bf N}(J)$.
The expression~(\ref{a1}) is reproduced if the source term has
two non--zero matrix elements, $\langle a (a+1) \mu | J | (a+n)
(a+n+1) \nu \rangle$ and $\langle (a+n) (a+n+1) \nu | J | a (a+1)
\mu \rangle$. The loop expansion generates contributions in which,
besides the $\delta \sigma^{(k)}$'s, powers of $J$ and $v$ appear
to all orders. We clearly need consider only terms of second order
in $J$. We first consider such terms which are of zeroth order in
$v$. Inspection shows that, for $n \gg 1$ and to arbitrary order
in the $\delta \sigma^{(k)}$'s, the result Eq.~(\ref{eq24}) remains
valid after Wick contraction of the $\delta \sigma^{(k)}$'s. Terms of
order $v^2$ come in two classes. There are corrections of order $v^2$
which either appear under the same trace(s) as the source terms to
begin with, or which become connected to the source terms through Wick
contractions of the $\delta \sigma$'s, and there are other such
corrections for which this is not the case. Formally, the latter can
be calculated by putting the source terms equal to zero. We can show
that after Wick contraction, every one of these latter terms takes the
form
\ba
&& c \ v^2 \sum_{j = 1}^{K'} {\rm trg} \biggl ( T^{(j)} L
(T^{(j)})^{-1}
\nonumber \\
&& \times T^{(j+1)} L (T^{(j+1)})^{-1} \biggr ) \ ,
\label{eq27}
\ea
with $c$ a constant which depends on the particular term considered.
Re--exponentiation of this result yields a renormalization of the
transport coefficient $T_{\rm sp}$. This is a very satisfying result.
Although we are not able to calculate the values of the coefficients
$c$ analytically, we can show that each of these coefficients is
determined entirely by two neighboring boxes, say $(aa)$ and
$(a(a+1))$. Any such pair of boxes yields the same result. Thus, our
analytical approach has reduced the calculation of the renormalized
transport coefficient to a problem involving only two neighboring
boxes, and not the full Hamiltonian $H$. To understand the physical
origin and meaning of the renormalization terms in this reduced
framework, we use the diagrammatic expansion of Ref.~\cite{agassi}
which is akin to a diagrammatic impurity perturbation expansion. One
can see heuristically that the correction terms add up to yield the
ensemble average of the product of the physical level densities
$\rho_0(E)$ and $\rho_1(E)$ in the two boxes. Thus, the sum of the
saddle--point value and of all the loop corrections yields for the
renormalized transport coefficient $T$ the expression
\be
T(w) = 2 \pi \ v^2 \ \overline{\rho_{0}(E, w) \rho_{1}(E)} \ .
\label{eq25}
\ee
The ensemble average in Eq.~(\ref{eq25}) can be determined without
difficulty by numerical simulation for every given two--body
interaction $w$. We must finally take the continuum limit by letting
the longitudinal length $d$ of each slice go to zero and the number
$K$ of slices go to infinity. In this limit we have
\be
\frac{\zeta(w \neq 0)}{\zeta(w = 0)} = {\rm lim}_{d \rightarrow
  0} \frac{T(w \neq 0)}{T(w = 0)}\ . 
\label{eq26}
\ee
We believe that in a numerical simulation, the ratio of localization
lengths will be fairly independent of $d$. 

{\it Summary}. We have studied the localization properties of two
interacting electrons in a quasi one--dimensional wire. Both electrons
move in a disorder potential and interact when they are close to each
other. Hopping allows the electrons to move along the wire.  
 
Our central result is contained in Eqs.~(\ref{eq25},\ref{eq26}). The
two--body interaction affects the localization length because it alters
the level density in the boxes containing two electrons. Depending on
the energy chosen and on the sign of the two--body interaction, the
localization length may decrease or increase. To the best of our
knowledge, this  is the first time that an analytical treatment of
both, disorder and  interaction, has been possible.

The supersymmetry method does not apply to one--dimensional systems.
Hence it is difficult to compare our results with previous numerical
work. The exception is the work of Imry~\cite{imry} who derives an
expression (his Eq.~(4)) for the change of the localization length
which applies also in higher dimensions and is proportional to
$(U/B)^2$, with $U$ a typical two--body matrix element and $B$ the
bandwidth. This expression differs from our expression~(\ref{eq25}) in
which we retain the structure typical for the Thouless block scaling
argument with hopping. We predict a change of localization length
which depends upon all moments ${\rm tr}(w), {\rm tr}(w^2), \ldots$
of the two--body interaction while Imry's result involves only his
$U^2$ and is, thus, independent of the sign of $w$.


\begin{thebibliography}{99}

\bibitem{dima}D. L. Shepelyansky, Phys. Rev. Lett. {\bf 73} (1994)
  2607; see also O. N. Dorokhov, Zh. Exp. Teor. Fiz. {\bf 98} (1990)
  646 [ Sov. Phys. JETP {\bf 71} (1990) 360 ]

\bibitem{imry}Y. Imry, Europhys. Lett. {\bf 30} (1995) 405
  
\bibitem{song}P. H. Song and F. von Oppen, Phys. Rev. {\bf 59} (1999)
  46

\bibitem{ben}L. Benet, T. Rupp, and H. A. Weidenm\"uller, Phys. Rev. 
    Lett. 87 (2001) 010601-1; Ann. Phys. (N.Y.) 292 (2001) 67

\bibitem{iid}S. Iida, H. A. Weidenm\"uller, and J. Zuk,
  Ann. Phys. (N.Y.) {\bf 200} (1990) 219

\bibitem{efe}K. B. Efetov, Adv. in Physics {\bf 32} (1983) 53

\bibitem{ver}J. J. M. Verbaarschot, M. R. Zirnbauer, and H. A.
Weidenm\"uller, Phys. Rep. {\bf 129} (1985) 367

\bibitem{agassi}D. Agassi, G. Mantzouranis, and H. A. Weidenm\"uller,
  Phys. Rep. {\bf 22} (1975) 145

\end{thebibliography}
\end{document}